# Maximum Mass of Hybrid Stars in the Quark Bag Model

**G. B. Alaverdyan**[*)] **and Yu. L. Vartanyan**

*Yerevan State University, A.Manoogyan str. 1, 0025 Yerevan, Armenia*

*The effect of model parameters in the equation of state for quark matter on the magnitude of the maximum mass of hybrid stars is examined. Quark matter is described in terms of the extended MIT bag model including corrections for one-gluon exchange. For nucleon matter in the range of densities correspond-ing to the phase transition, a relativistic equation of state is used that is calculated with two-particle correlations taken into account based on using the Bonn meson-exchange potential. The Maxwell construction is used to calculate the characteristics of the first order phase transition and it is shown that for a fixed value of the strong interaction constant $\alpha_s$, the baryon concentrations of the coexisting phases grow monotonically as the bag constant B increases. It is shown that for a fixed value of the strong interaction constant $\alpha_s$, the maximum mass of a hybrid star increases as the bag constant B decreases. For a given value of the bag parameter B, the maximum mass rises as the strong interaction constant $\alpha_s$ increases. It is shown that the configurations of hybrid stars with maximum masses equal to or exceeding the mass of the currently known most massive pulsar are possible for values of the strong interaction constant $\alpha_s > 0.6$ and sufficiently low values of the bag constant.*

## 1. Introduction

The question of the possible presence of quark matter in the depths of compact stars has been at the center of attention of theoretical physicists for the last few decades. During this time, many different equations of state for

---

[*)] e-mail: galaverdyan@ysu.am

superdense matter have been constructed which allow a phase transition of matter from a hadronic composition to matter made up of deconfined quarks. The integral and structural characteristics of neutron stars, whose central part consists of a quark-electron *udse* plasma, have been calculated in terms of these models. Compact stars with this kind of structure are referred to a hybrid stars. The detection of massive millisecond pulsars with masses on the order of twice the sun's mass (PSRJ1614-2230, with a mass of $1.97 \pm 0.04\, M_\odot$ [1] and subsequently refined value of $1.928 \pm 0.017\, M_\odot$ [2], and PSRJ0348+0432, with a mass of $2.01 \pm 0.04\, M_\odot$ [3]) has stimulated new work [4-8]. The main purpose of these papers has been to develop models for an equation of state of superdense matter that can yield the large masses of compact stars containing quark matter. It is known that the phase transition to quark matter leads to a softening of the equation of state and, thereby, to a reduction in the maximum mass of a stellar configuration. In this regard it is necessary to clarify how much this softening in a given model affects the maximum mass of a star and how this conforms with the observational limit owing to the existence of the above-mentioned massive pulsars.

In this paper we study hybrid stars under the assumption that the surface tension between hadrons and quark matter is so strong that the hadron-quark phase transition satisfies Maxwell's conditions, which lead to a discontinuous change in the density of the substance under pressures corresponding to thermodynamic equilibrium between the two phases. For hadron matter at densities near and above nuclear, a relativistic equation of state is used which is calculated with two-particle correlations taken into account using the Bonn meson-exchange potential [9]. Quark matter is described by an equation of state calculated in terms of the MIT bag model [10] containing first order correction terms with respect to the strong interaction constant.

## 2. Equation of state of matter with a hadron-quark phase transition

The equation of state of matter with a hadronic structure is constructed in this paper by matching four equations of state. In the region of densities below the saturation density for nuclear matter, the Feynman-Metropolis-Teller [12], Baym-Pethick-Sutherland [13], and Baym-Bethe-Pethick [14] equations are used. In the near-nuclear and above-nuclear density regions a relativistic equation of state is used which takes two-particle correlations into account based on application of the Bonn meson-exchange potential [9].

Three-flavor quark matter consisting of *u*, *d*, and *s* quarks and electrons is described using an extended MIT quark bag model in which the interactions between quarks inside the bag are accounted for in the approximation of one-gluon exchange [11]. The thermodynamic properties of quark matter are derived from the grand thermodynamic potentials $\Omega_i$ ($i = u, d, s, e$) per unit volume. The thermodynamic potential of a quark component depends both on the chemical potentials and on the masses of the quarks, the phenomenological parameter $B$ of the bag model, and the quark-gluon interaction constant $\alpha_s$:

$$\Omega_Q(\{\mu_f\}, \{m_f\}, B, \alpha_s) = \sum_{f=u,d,s} \Omega_f(\mu_f, m_f, \alpha_s) + B. \qquad (1)$$

The equation of state for a β-equilibrium electrically neutral quark-electron plasma is defined in parametric form as

$$P(\mu_B^{QM}) = -\sum_{i=u,d,s,e} \Omega_i - B,$$
$$\varepsilon(\mu_B^{QM}) = \sum_{i=u,d,s,e} (\Omega_i + \mu_i n_i) + B, \quad (2)$$
$$n^{QM}(\mu_B^{QM}) = \frac{1}{3}(n_u + n_d + n_s),$$

where $\mu_B^{QM}$ is the baryon chemical potential and $n^{QM}$ is the baryon number density of the quark phase.

For the current masses of the quarks we used the values $m_u=3$MeV, $m_d=5$MeV, and $m_s=100$MeV in accordance with the latest data [15]. The numerical values of the model, $B$ and $\alpha_s$, are not known exactly. Since the expressions for the thermodynamic potentials $\Omega_f$ of the quarks also include a first order term in the strong interaction constant $\alpha_s$, in our calculations we limited ourselves to variations over the range $0 \leq \alpha_s \leq 0.4$. To determine the possible values of the bag constant $B$ (for given $\alpha_s$) corresponding to a first order phase transition between matter consisting of hadrons and quark matter consisting of free quarks $u$, $d$, $s$, and electrons, we proceeded from the following arguments in accordance with Refs. 11 and 16. Hybrid stars will exist if the model parameters ensure the coexistence of hadron and quark phases. In the Maxwell construction it is assumed that each phase separately is electrically neutral and at some pressure $P_0$ the baryon chemical potentials of both phases approach one another:

$$P^{(HM)} = P^{(QM)} = P_0, \quad \mu_B^{(HM)}(P_0) = \mu_B^{(QM)}(P_0). \quad (3)$$

The hypothesis that three-flavor *uds* matter can be absolutely stable, i.e., be the ground state of the matter at zero pressure, will hold if the energy per baryon at zero pressure is lower than the analogous characteristic of the most bound iron nucleus, i.e.,

$$E_1^{uds} < \frac{M_{^{56}Fe}}{56} = 930.4 \text{ MeV}$$

At the same time, the energy per baryon for the two-color electrically neutral, non-strange *ud* matter cannot be less than the neutron rest energy or else the neutrons would be depleted of the larger droplets of *ud* matter, i.e., $E_1^{ud} > 939.6$ MeV.

The equation

$$E_1^{uds}(B, \alpha_s) = 930.4 \quad (4)$$

determines the upper bound of the parameter $B_{\max}^{SS}(\alpha_s)$ corresponding to the existence of self-bound *uds* matter and,

therefore, of such stellar objects as strange stars. The upper bound $B_{max}^{SS}(\alpha_s)$ for strange stars is simultaneously the lower bound of the bag parameter for quark-hadron hybrid stars $B_{min}^{QHS}(\alpha_s) = B_{max}^{SS}(\alpha_s)$.

The equation

$$E_1^{ud}(B, \alpha_s) = 939.6 \tag{5}$$

determines the lower bound of the parameter $B_{min}^{SS}(\alpha_s)$ corresponding to the existence of strange stars.

Figure 1 shows the regions corresponding to strange and hybrid stars in the $B - \alpha_s$ plane. The region bounded by the curves $B_{min}^{SS}(\alpha_s)$ and $B_{max}^{SS}(\alpha_s)$ corresponds to combinations of the bag model parameter $B$ and $\alpha_s$ that lead to the existence of strange stars. The region above the curve $B_{max}^{SS}(\alpha_s)$ corresponds to hybrid quark stars. The combination of $B$ and $\alpha_s$ in this region ensures that for some value of the pressure $P_0$, the conditions (3) for the existence of two phases will be met. In order to illustrate the sensitivity of the boundary $B_{max}^{SS}(\alpha_s)$ to the strange quark mass, curves are shown on the graph that correspond to the strange quark masses, $m_s = 95, 100, 120, 140,$ and

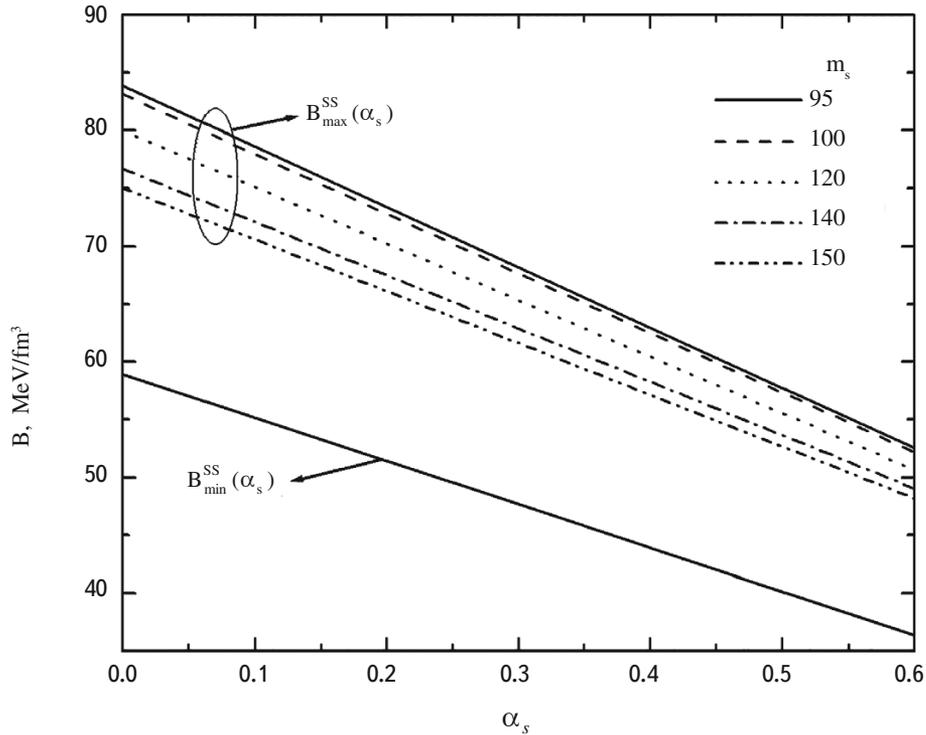

Fig. 1. The ranges of values of the bag model parameter $B$ and $\alpha_s$ corresponding to strange (SS) and hadron-quark hybrid (HQS) stars. The region between the curves $B_{min}^{SS}(\alpha_s)$ and $B_{max}^{SS}(\alpha_s)$ corresponds to strange stars. The region above the curve $B_{max}^{SS}(\alpha_s)$ corresponds to hybrid stars.

150 MeV.

It is known that for a Maxwellian phase transition, the baryon number density $n$ and energy density $\varepsilon$ have a discontinuity at $P_0$. The baryon number density at the transition point for the nucleon phase is given by $n_N = n^{HM}(P_0)$ and for the quark matter, by $n_Q = n^{QM}(P_0)$. Figure 2 shows the threshold values of the baryon number densities $n_N$ and $n_Q$ as functions of the bag parameter $B$ for three values of the strong interaction constant $\alpha_s = 0$, 0.2, and 0.4. It can be seen that with increasing $B$, the concentrations for the transition grow monotonically. As the strong interaction constant $\alpha_s$ increases, the curves change slightly in shape and simultaneously shift toward lower values of $B$. Note that for small values of $B$ the baryon number density of the nucleon phase $n_N$ corresponding to phase equilibrium may be lower than the normal nuclear density $n_0 = 0.16$ fm$^{-3}$. We have encountered phase transitions of this type in our earlier work [17,18] in which the characteristics of a Maxwellian phase transition were calculated and neutron stars with cores of quark-electron matter were studied by comparing several variants of the equations of state for nucleon matter with variants of the equation of state for quark matter obtained using the MIT bag model for different values of the model parameters $m_s$, $B$, and $\alpha_s$.

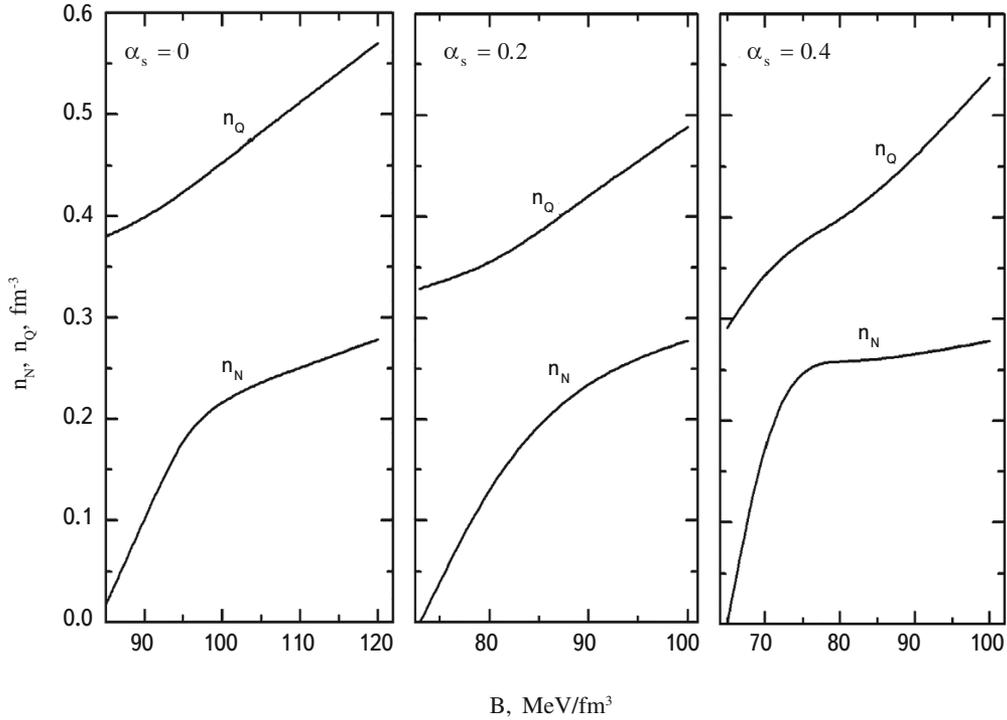

Fig. 2. The threshold values of the baryon concentrations $n_N$ and $n_Q$ of the phase transition as functions of the bag parameter B for three values of the strong interaction constant $\alpha_s = 0$, 0.2, and 0.4.

## 3. Configurations of hybrid stars with maximal mass

Using a set of different values of the parameters $B$ and $\alpha_s$ and MIT bag models corresponding to the Maxwellian scenario for a first order phase transition, we have integrated the Tolman-Oppenheimer-Volkoff system of equations

$$\frac{dP}{dr} = -\frac{G}{r^2 c^2} \frac{(\varepsilon(P)+P)}{1-\frac{2Gm}{rc^2}} \left( m + \frac{4\pi r^3 P}{c^2} \right), \quad \frac{dm}{dr} = \frac{4\pi r^2}{c^2} \varepsilon(P) \qquad (6)$$

for different values of the central pressure and found the parameters of hybrid star configurations with maximum masses. Table 1 lists the parameters of the configurations with maximum mass for four sets of quark bag parameters.

Figure 3 shows the contours of the function $M_{max}(B,\alpha_s)$ in the $B-\alpha_s$ plane for different values of the maximum mass of hybrid stars. This figure shows that the maximum mass is relatively higher for small bag constants $B$ and large strong interaction coupling constants $\alpha_s$. But for fixed values of $\alpha_s$, the constant $B$ is bounded below by $B_{min}^{QHS}(\alpha_s) = B_{max}^{SS}(\alpha_s)$ derived from Eq. (4). For a given value of $\alpha_s$, this limit on $B$ also leads to an upper bound on the maximum mass. As $\alpha_s$ increases, the lower bound on the permissible values of $B$ decreases; this, in turn leads to an increase in the upper bound for the maximum mass of hybrid Maxwellian stars. By increasing $\alpha_s$ and simultaneously reducing $B$ to $B_{min}^{QHS}(\alpha_s)$, of course, it is possible to attain maximum masses of hybrid stars in excess of $2 M_\odot$. But it should be noted that the expression for the thermodynamic potential based on the extended bag model [11] includes a correction for interactions among quarks inside the bag in a linear approximation in $\alpha_s$, which corresponds to one-gluon exchange. This leads to the question of whether the use of the extended MIT bag model is permissible for large values of the strong interaction constant $\alpha_s$. In addition, when $B$ is reduced, as noted above there is a reduction in the baryon number density of the nucleon phase, $n_N = n^{HM}(P_0)$, corresponding to the phase equilibrium. For sufficiently low $B$, the density $n_N$ becomes so much lower than the normal nuclear density that the question of the irreality of this choice of values for the model parameters $B$ and $\alpha_s$ arises.

TABLE 1. The Parameters of Series of Hybrid Stars with Maximum Mass. Radius $R_m$ and Central Pressure $P_c^{(m)}$ Corresponding to the Configuration with Maximum mass $M_{max}$.

| $\alpha_s$ | B, MeV/fm³ | $M_{max}/M_\odot$ | $R_m$, km | $P_c^{(m)}$, MeV/fm³ |
|---|---|---|---|---|
| 0.2 | 73 | 1.72 | 9.43 | 400 |
|  | 90 | 1.57 | 9.38 | 512 |
| 0.4 | 65 | 1.82 | 10.01 | 350 |
|  | 80 | 1.68 | 10.22 | 394 |

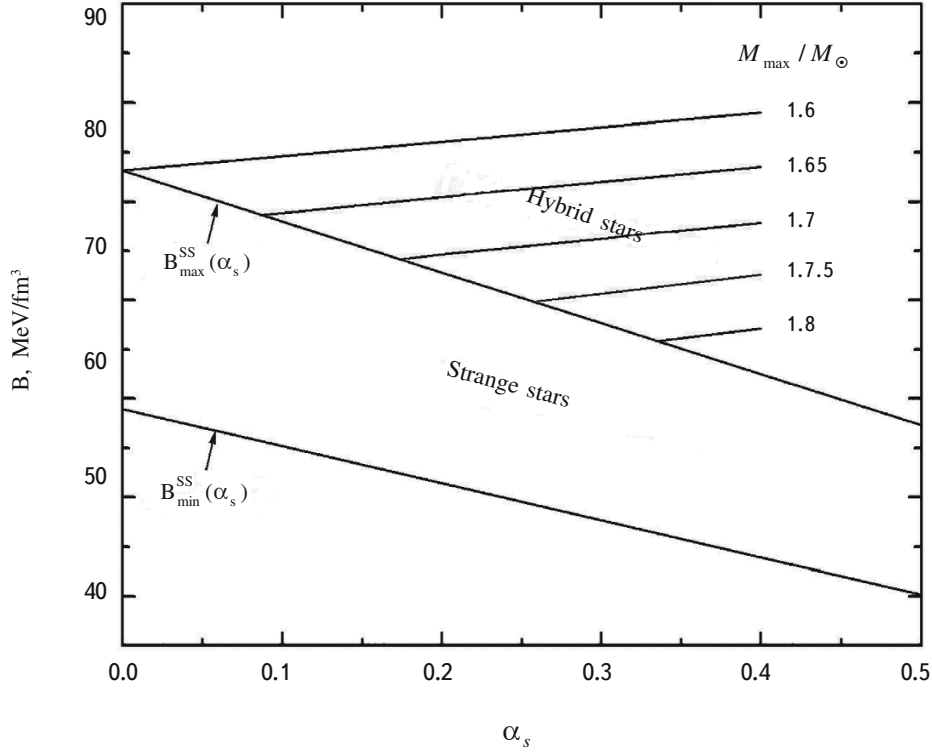

Fig. 3. Contours of the function $M_{max}(B, \alpha_s)$ for different values of the maximum mass of hybrid stars.

## 4. Conclusion

In this paper we have studied quark-hadron hybrid stars using an equation of state derived from the extended MIT bag model. For nucleon matter in the near-nuclear density range, a relativistic equation of state has been used that includes two-particle correlations based on the Bonn meson-exchange potential. Ranges of the bag model parameters $B$ and $\alpha_s$ have been found which correspond to hybrid stars, as well as to self-bound three-flavor quark matter and strange stars. For different combinations of $B$ and $\alpha_s$ belonging to the hybrid stars region, a Maxwell construction has been used to calculate the characteristics of the first order phase transition and it has been shown that for a fixed value of the strong interaction constant $\alpha_s$, the baryon concentrations of the coexisting phases grow monotonically with increases in the bag constant $B$. For low values of $B$, the baryon number density $n_N$ of the nucleon phase corresponding to the phase equilibrium can take values below the normal nuclear density $n_0$. The integral parameters of hybrid stars have been calculated by numerical integration of the system of TOV equations and the maximum mass has been found as a function of the model parameters $B$ and $\alpha_s$. It has been shown that for a given value of $\alpha_s$ the maximum mass of a hybrid star increases as the bag constant $B$ is reduced. For fixed $B$, the maximum mass increases as $\alpha_s$ rises.

Our analysis in terms of the extended MIT bag model shows that configurations of hybrid stars with maximum masses equal to or exceeding the mass of the currently known most massive pulsar are possible for strong interaction constants $\alpha_s > 0.6$ and sufficiently low values of the bag constant close to the value $B_{\min}^{QHS}(\alpha_s)$ given by the condition (4). But the applicability of the one-gluon exchange approximation at these values of $\alpha_s$ remains an open question.

Since the properties of hybrid stars depend on the equations of state of hadron as well as quark matter, a complete analysis of and reliable conclusions regarding the possible presence or absence of quark matter in the depths of compact stars will require examination of a combination of relativistic equations of state for hadron matter and the equation of state of quark matter. A separate paper will be devoted to studies of configurations of hybrid stars with maximum mass employing these kinds of combinations of equations of state, where hadron matter is described in terms of a relativistic mean field theory and quark matter, in terms of the extended MIT bag model.

This work was done in the scientific-research laboratory for the physics of superdense stars at the department of the theory of wave processes and physics of Erevan State University, which is supported by the State committee of science of the Ministry of education and science of the Republic of Armenia.